\documentclass[12pt]{iopart}
\newcommand{\rvac}{\rho_{\rm vac}}

\newcommand{\vr}{\textbf{r}}

\newcommand{\lp}{\left(}
\newcommand{\rp}{\right)}

\newcommand{\dd}{{\rm d}} 
\newcommand{\dtr}{\dd^{3}r} 

\newcommand{\om}{\mathbf{\Omega}}

\newcommand{\maw}{\mathcal{W}}

\newcommand{\mai}{\mathcal{I}}

\newcommand{\mar}{\mathcal{R}}

\newcommand{\mapp}{\mathcal{P}}
\newcommand{\maa}{\mathcal{A}}

\usepackage{graphicx}

\begin{document}
\title{Ellipsoidal configurations in the de Sitter spacetime}

\author{A. Balaguera-Antol\'{\i}nez}
\address{Departamento de F\'{\i}sica, Universidad de los Andes,A.A. 4976, Bogot\'a, D.C., Colombia.}
\ead{a-balagu@uniandes.edu.co}
\author{D. F. Mota}
\address{Institute of Theoretical Astrophysics, University of Oslo, 0315 Oslo, Norway}
\address{Perimeter Institute for Theoretical Physics, Waterloo, Ontario N2L 2Y5, Canada}
\ead{d.f.mota@astro.uio.no}
\author{M. Nowakowski}
\address{Departamento de F\'{\i}sica, Universidad de los Andes,A.A. 4976, Bogot\'a, D.C., Colombia.}
\ead{ mnowakos@uniandes.edu.co}

\begin{abstract} 
The cosmological constant $\Lambda$ modifies certain properties of large astrophysical rotating  
configurations with ellipsoidal geometries, provided the objects are not too compact.   
Assuming an equilibrium configuration and so
using the tensor virial equation with $\Lambda$ we explore several equilibrium 
properties of homogeneous rotating ellipsoids.
One shows that the bifurcation point, which in the oblate case distinguishes 
the Maclaurin ellipsoid from the Jacobi ellipsoid, is sensitive to the cosmological constant.  
Adding to that, the cosmological constant allows triaxial configurations  
of equilibrium rotating the minor axis as solutions of the virial equations.  
The significance of the result lies in the fact that minor axis rotation 
is indeed found in nature. Being impossible for the oblate case, it is permissible  
for prolate geometries, with $\Lambda$ zero and positive. 
For the triaxial case, however, an equilibrium solution is found only for non-zero positive $\Lambda$. 
Finally, we solve the tensor virial equation for the angular velocity and display special effects of 
the cosmological constant there.  
\end{abstract}


\date{}
\maketitle

\section{Introduction} 
 
In the last decades the universe has been mapped by different methods sufficiently accurate 
to allow us global and detailed conclusions about its properties and behavior.  
Key observations are the luminosity-redshift relationship of supernovae of  
type Ia (SNIa) \cite{Riess}, the matter power spectrum of large scale structure as inferred  
from galaxy redshift surveys \cite{sloan},  
and the anisotropies in the Cosmic Microwave Background Radiation \cite{wmap}.  
These observations allows us to conclude that the Universe is spatially flat, and surprisingly  
its expansion is accelerating (instead of the long thought slowing down). 
Many different models have been proposed to explain this acceleration: For instance,
large scale modifications of General Relativity \cite{hog1}. The most popular explanation is, however, the inclusion of a dark energy component 
which corresponds to about $70\%$ of the total energy budget of the universe and has a quite negative pressure. Candidates for the dark energy range from 
scalar fields \cite{dde1} (which are widely used in cosmology \cite{scalar}), up to  dark energy/dark matter unifying candidates, as for instance the Chaplygin Gas \cite{cg1}. 
The simplest model, however, would still be a positive cosmological constant $\Lambda$. 
%

Understanding the nature of dark energy is one of the major objectives in cosmology nowadays. 
One way to undisclose its properties is to investigate its astrophysical and cosmological imprints.
Dark energy affects the formation of galaxies and clusters \cite{wang, nunes}, as well as dark matter halo properties, such as the virial mass 
and the virial radius \cite{mota}.  In \cite{MNS} it was also shown that $\Lambda$ affects the motion of test bodies 
in the Schwarzschild-de Sitter metric over large distances, a case of importance for large scale objects.
Although observables within the solar system seem not to be sensitive to the global expansion of the universe \cite{Iorio}, that is not the case with
 large small-density astrophysical objects. In fact, investigations on the effects of dark energy in the local Hubble flow show that 
those effects may not be negligible \cite{new2}. 
In case of a cosmological constant, it has been shown that not only the formation
of astrophysical bodies gets affected by $\Lambda$, but also its static properties like the internal mean velocity 
and the angular velocity \cite{MNCh}. In fact, it was shown in \cite{new1} that 
$\Lambda$ may produce lower values of galaxy velocity dispersions. 
Taken all these findings together one can state that dark energy has some effects at astrophysical scales. 
It is then important to try to understand what further possible effects could dark energy imprint on some of the properties of the astrophysical bodies. 
These signatures can be used to  probe and test dark energy candidates.


The effect of $\Lambda$ is bigger the more we deviate 
the object's geometry from the spherical case \cite{MNS,MNCh}
. This is important since 
by far not all structures found in the universe have spherical symmetry.  
Rotating configurations acquire non spherical geometries which arise as solutions of the tensor virial theorem (of second order).  
With zero cosmological constant some of these figures of equilibrium correspond to Jacobi ellipsoid (triaxial, rotating along the minor axis)  
and Maclaurin ellipsoid (oblate, rotating along the minor axis). Also one finds the Dedekin ellipsoid,  
(triaxial and flat because internal streaming), the Riemann ellipsoid (triaxial with rotation and internal motion),  
among other possible solutions. 
Interestingly, not only galaxies have different morphologies \cite{morph}, but also 
clusters of galaxies. Of interest for us is especially the fact that clusters can have a prolate shape \cite{prolatecl} (hence we expect that they 
can also appear in triaxial configurations) and they can rotate \cite{rot}. 

In this article we will investigate some basic properties of ellipsoids in the presence of a cosmological constant. In particular, we will
answer questions related to
the geometry of a rotating configuration. For instance, given the axis 
of rotation, which could be any one of the three axis, what is the allowed range of the axis which emerges as a solution? 
A legitimate question is also  
what are the changes regarding the bifurcation points? In other words, 
at which point is a triaxial solution allowed. Finally, we can ask what is the possible angular momentum? 
Although not frequently encountered, there are astrophysical objects rotating along the major axis (this is the same to say that 
the objects rotates the minor axis) \cite{minoraxis1,minoraxis3a}. 
It is interesting to make some theoretical attempts to understand 
such exotic configurations. We will show that the triaxial case \cite{binneyart1} of minor axis rotation is possible with positive $\Lambda$  
and the allowed region associated to values characterizing the geometrical configuration vanishes with vanishing $\Lambda$. 
For oblate and prolate case we calculate the bifurcation points emphasizing the modifications and 
new effects induced by $\Lambda$. 
Clearly, the ellipsoidal geometry which we will study 
in the paper is well suited to model many geometries such as the disk. Therefore, the results of this article should in principle be 
applicable also to superclusters which are usually found to be very flat objects \cite{supercl}.

Hereby we will concentrate on homogeneous rotating ellipsoids,  
whose equilibrium configuration will be explored by  
the tensor virial equation with $\Lambda$ \cite{barrow,MN} describing the equilibrium properties of  
possible non-spherical configurations. 
The tensor virial theorem results into two equations: an equation for the angular velocity and one 
restricting possible geometries. It is therefore a perfect tool to study the subject. 
It is known that for spherical systems it is sufficient to work with the scalar virial theorem. 
Our results generalize the ones obtained in \cite{Chandra}. This is especially true for the 
bifurcation points of rotating ellipsoids as solutions of tensor virial equations for the oblate geometry. 
Indeed, we follow closely the exposition and methods given in \cite{Chandra} which applies for the case 
$\Lambda=0$. However, the study of the prolate/triaxial case with minor axis rotation is new.  It is not surprising after all that 
this has not been examined before as this case is only possible with $\Lambda \neq 0$.

This paper is organized as follows: in section 2, we review the basic expressions for  
describing equilibrium in the Newtonian limit with a cosmological constant. Since we retain 
in this limit $\Lambda$, we are essentially dealing with a non-relativistic de Sitter spacetime (also called Newton-Hooke).  
Section 3 is devoted to a brief discussion about the applicability of our approach to
galaxy clusters.
In section 4 we explore the allowed configurations for a rotating ellipsoid whose rotation lies  
along the minor axis (major axis rotation).  
The next section explores the strange scenario,  
which corresponds to the case of minor axis rotation, i.e. when the angular velocity is along the major axis. We end with conclusions.

\section{The Newtonian limit with cosmological constant} 
 
To explore the consequences of a positive cosmological constant, we make use of 
the Newtonian limit of Einstein field equations. In this limit, the Poisson equation is written as  
\begin{equation} 
\label{poisson} 
\nabla^{2}\Phi=4\pi G_{N}\lp \rho+3\frac{P}{c^{2}}\rp - \Lambda. 
\end{equation} 
where $P$ is the pressure, which arises as a general relativity correction.  
The solution of (\ref{poisson}) at the zeroth order of $v/c$ (from now on we set $c=1$) takes the form 
\begin{equation} 
\label{psolution} 
\Phi(\vr)=-G_{N}\int_{V'}\frac{\rho(\vr')}{|\vr-\vr'|}\,\dd ^{3}r'-\frac{1}{6}\Lambda |\vr|^{2}, 
\end{equation} 
where we have neglected secondary effects of the cosmological constant in the set of the Dirichlet boundary conditions \cite{MNL}.  
The potential is then a contribution from a pure gravitational part $\Phi^{\rm grav}$  
and a contribution from the expansion proportional to $\Lambda$. 
 
Together with equation (\ref{poisson}), a self gravitating configuration  
is ruled by an equation of state and the hydrostatic equilibrium equation (Euler's equation), which is written as  
\begin{equation} 
\label{euler} 
\rho \frac{\dd u_{i} }{\dd t}+\rho \partial_{i}\Phi+\partial_{j}\mapp_{ij}=0, 
\end{equation} 
where $\mapp_{ij}$ is the pressure tensor (which for simplicity we take as isotropic $\mapp_{ij}=\delta _{ij}P$)  
and $u_{i}=\langle v_{i}\rangle$ is the mean values of the velocity.  
The second order tensor virial equation can be derived by taking spatial moments in Euler's equation.  
The result is an integral equation 
\begin{equation} 
\label{virotro} 
\frac{1}{2}\frac{\dd ^{2}\mai_{ik}}{\dd t^{2}}=2T_{ik}-|\maw_{ik}| 
+\frac{8}{3}\pi G_N \rho_{\rm vac}\mai_{ik}+ \Pi_{ik}, 
\end{equation}  
with  
\begin{eqnarray} 
\label{defi} 
\Pi_{ik}=\int_{V}\mapp_{ik} \dtr,\hspace{0.2cm}T_{ik}\equiv \frac{1}{2}\int_{V}\rho \langle v_{i}\rangle \langle v_{k}\rangle \dtr, \\ 
\maw_{ik}=-G_{N}\int_{V}\rho(\vr)r_{i}\partial_{k} \Phi^{\rm grav}\dtr. 
\end{eqnarray}  
which corresponds to the pressure integral, the kinetic energy tensor and the gravitational potential  
energy tensor, respectively.  
The moment of inertia tensor is defined as  
\begin{equation} \label{inertia13} 
\mai_{ik}\equiv \int_{V}\rho r_{i}r_{k}\dtr.  
\end{equation} 

In statistical mechanics  
the virial theorem follows from taking time averages over a long period of time (denoted below by $\langle...\rangle$) 
such that the configuration reaches equilibrium.  
This implies that in equilibrium, the left hand side of (\ref{virotro}) vanishes which gives  
us the tensor virial theorem as 
\begin{equation} 
\label{tvt} 
2\langle T_{ik}\rangle_{\rm T}-\langle|\maw_{ik}|\rangle_{\rm T} 
+\frac{8}{3}\pi G_N \rho_{\rm vac}\langle\mai_{ik}\rangle_{\rm T}+ \langle\Pi_{ik}\rangle_{\rm T}=0. 
\end{equation}  
In astrophysics this would imply that we have to take an average over ensembles of observed objects of a 
specific type. This, in principle, is possible 
noting that the fact that the external contribution, i.e, the vacuum energy density $\rho_{\rm vac}\equiv \Lambda/8\pi$ is constant,  
and therefore the time average is only taken over proper parameters of the configuration. 
However, it is also customary and more convenient to assume that for most astrophysical objects 
the second derivative of the inertial tensor vanished at a sufficiently large time. The averaging is not necessary then.  
For generalizations (dark energy , scalar fields or Chaplygin gas), where one has also to take into account  
the evolution of the background, this is a more straightforward approach.  
 
In the next sections, we drop the $\langle...\rangle$ brackets and also use the convention $G_{N}=c=1$. 

\section{The virial theorem and large astrophysical structures}

It is of some importance to this work to discuss briefly the question whether the virial
theorem or hydrostatic equilibrium apply to large astrophysical structures.
This is of importance  for the present study 
because certainly the effect of $\Lambda$ will be seen only for relatively low density objects, i.e. clusters
and superclusters. The other issues which are of phenomenological interest for us 
is the shape of these objects and their possible rotation.

Whether one can apply the virial theorem (or hydrostatic equilibrium) and its consequences to 
groups or clusters of galaxies has been a matter of debate over years \cite{smith, biviano1, biviano2}.
In many cases (and these are the cases to which we can apply our subsequent results), this dispute has
been decided in favour of the virial theorem \cite{biviano1, postman}. Indeed, over decades the virial
technique is the most widely used one \cite{turner, bahcall} and the agreement with other methods is satisfactory
\cite{rhines}.  The assumption of hydrostatic equilibrium in clusters is common and, as it seems, works well
\cite{reipich}. There are exceptions due to cooling flows \cite{sodre} or misidentification of the cluster
shape (i.e. superposition of two apparent individual objects) \cite{rhines2}. 
Last but not least, the virial theorem
with cosmological constant has been successfully applied to galaxy clusters at early epochs \cite{wang}.  

The Morphology of clusters is as rich as the corresponding one for galaxies \cite{morph}. Almost all
shapes can be fitted into ellipsoids \cite{prolatecl, rot, novikov, binggeli}. 
Interestingly, virial results are used to take into account finite size effects of galaxy clusters
in lensing \cite{oguri}. These effects are parameterized by ellipticity, a concept related to eccentricity of ellipsoids.
The second important feature of clusters, mentioned already in the introduction, is that
they can exhibit rotation \cite{rot} exactly as galaxies do. The applicability of virial theorem to clusters taken together
with the facts of their ellipsoidal nature and the established rotation means that we can apply our results below to
galaxy clusters. It remains to address the question whether the application is valid for the luminous matter
or the `whole' object including dark matter. We can rely here on studies and simulations which all, and this is the
crucial observation here, indicate that dark matter halos around clusters are of ellipsoidal nature \cite{floor, paz}.
However, we could apply the virial theorem to the luminous structure alone including for the total mass also the
dark matter component. This seems possible as the form of the luminous matter does not change over
cosmological times and hence can be assumed to be in equilibrium as a substructure. In any case, the
morphology under discussion here will be ellipsoidal.  

Going from clusters to superclusters the following observation is worth noticing. Clusters are often elongated within their group
\cite{novikov} and aligned within their supercluster \cite{binggeli}. 
This filamentary network \cite{zeldovich} (.e.g. between the Virgo and
A1367 cluster \cite{ramella}) could be a virialized substructure (including the clusters themselves) of the supercluster.
Indeed, distinction between virialized and non-virialized regions 
of a supercluster exits \cite{reisenegger}. In such a case these substructure would be best described
by a prolate ellipsoid. If the angular momenta of the clusters are correlated \cite{aryal} we could have a model of 
rotating prolate/triaxial ellipsoid of clusters.

Finally, even though superclusters are, for different reasons, not believed to be in equilibrium \cite{bahcall, MN, Gramann},
the assumption of hydrostatic equilibrium still is often made in literature, either to test its grounds or to have a definite model
\cite{rhines2}. Even in cases where hydrostatic equilibrium is shown to be not valid, the results form the virial theorem seem
to be still useful \cite{rivolo} when applied to superclusters. This conclusion is corroborated in \cite{small}.

\section{Major axis rotation: triaxial and oblate ellipsoids} 
In this section we will consider the commonly known kind on ellipsoids configuration  
associated to cosmological and astrophysical systems, namely, oblate and triaxial configurations  
whose rotation is along the minor axis.  
Certainly a great portion of the number of cosmological structures have a non spherical symmetry,  
which can be related to internal streaming or rotation.  
 
Let us consider configurations with constant angular velocity along the minor axis.  
The kinetic energy tensor is then a contribution from internal motions and the rotating kinetic energy tensor, i.e.   
\begin{equation} 
T_{ik}\to T_{ik}+\mar_{ik}, 
\end{equation}  
where $\mar_{ik}$ is defined for constant angular velocity as 
\begin{equation}  
\label{rotation} 
\mar_{ik}\equiv \frac{1}{2}\lp\Omega_{\rm rot}^{2}\mai_{ik} -\Omega_{{\rm rot} i} 
\mai_{kj}\Omega_{j \rm rot}\rp, 
\end{equation} 
In the following analysis, we will neglect internal motion.  
The tensor virial equation (\ref{virotro}) for an homogeneous rotating configuration  
with angular velocity $\om=\Omega\hat{e}_{\rm z}$ axis is then 
\begin{equation} 
\label{vt} 
\Omega^{2}_{\rm rot}\lp \mai_{ik}-\delta_{iz}\mai_{iz}\rp-|\maw_{ik}|+\frac{8}{3}\pi\rho_{\rm vac}\mai_{ik}=-\delta_{ik}\int_{V}P\,\dtr, 
\end{equation}  
where $\maw_{ik}$ is the gravitational potential energy tensor and $\mai_{ik}$ is the moment of inertia tensor.  
From this expression one can derive two relevant equations.  
In the first place, we find the equation for the angular velocity, given by 
\begin{equation} 
\label{omega} 
\Omega^{2}_{\rm rot}=\frac{|\maw_{xx}|-|\maw_{zz}|}{\mai_{xx}}+\frac{8}{3}\pi\rho_{\rm vac}\lp\frac{\mai_{zz}}{\mai_{xx}}-1\rp, 
\end{equation}  
which reduces to the standard Maclaurin formula (in the oblate case) for homogeneous configurations putting  
$\rho_{\rm vac}=0$. On the other hand, by eliminating $\Omega^{2}$ from Eq.(\ref{vt}),  
we obtain a geometrical relation which restricts the geometry allowed for equilibrium configurations: 
\begin{eqnarray} 
\label{geom} 
\mai_{yy}\lp|\maw_{xx}|-|\maw_{zz}|\rp-\mai_{xx}\lp|\maw_{yy}|-|\maw_{zz}|\rp=\\\frac{8}{3}\pi \rho_{\rm vac}\mai_{zz}\lp\mai_{xx}-\mai_{yy}\rp \nonumber. 
\end{eqnarray}  
 
\begin{figure} 
\begin{center} 
\includegraphics[angle=270,width=8.cm]{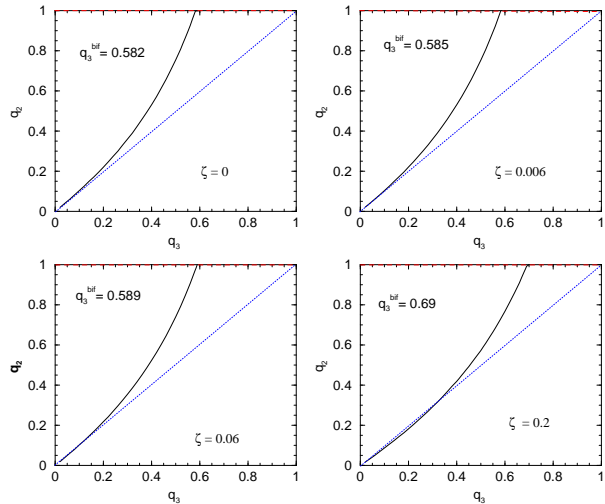} 
\end{center} 
\caption[Eccentricities II]{\footnotesize{Bifurcation Oblate-Triaxial: Values of $q_{2}$ and $q_{3}$ satisfying the virial  
condition condition $F(q_{2},q_{3})=0$.}}\label{bif2} 
\end{figure}

 
%
For a triaxial ellipsoid with constant density $\rho$ and semi-axis  
$a_{1}$, $a_{2}$ and $a_{3}$ with  
\begin{equation} \label{axis} 
a_{1}>a_{2}>a_{3},  
\end{equation} 
Eq. (\ref{geom}) can be recast into the following form 
\begin{equation} 
\label{gres} 
q_{2}^{2}=\frac{q_{3}^{2}(A_{3}(q_{2},q_{3})-\frac{2}{3}\zeta)}{A_{2}(q_{2},q_{3})-A_{1}(q_{2},q_{3})+q_{3}^{2}(A_{3}(q_{2},q_{3})-\frac{2}{3}\zeta)}, 
\end{equation} 
where  
\begin{equation} 
\zeta\equiv 2\rho_{\rm vac}/\rho,\hspace{1.5cm} q_{i}\equiv a_{i}/a_{1},  
\end{equation} 
with $q_{2}>q_{3}$. The functions $A_{i}$ are defined in dependence of $q_2$ and $q_3$ as (see \cite{Chandra,Roberts,Binney}) 
\begin{eqnarray} 
\label{a} 
A_{1}(q_{2},q_{3})&=&2q_{2}q_{3}\lp\frac{F(\theta,k)-E(\theta,k)}{k^{2}\sin^{3}\theta}\rp\\ \nonumber 
A_{2}(q_{2},q_{3})&=&2q_{2}q_{3}\lp\frac{E(\theta,k)-(1-k^{2})F(\theta,k)-\frac{q_{3}}{q_{2}}k^{2}\sin\theta}{(1-k^{2})k^{2}\sin^{3}\theta}\rp\\ \nonumber 
A_{3}(q_{2},q_{3})&=&2q_{2}q_{3}\lp\frac{\frac{q_{3}}{q_{2}}\sin\theta -E(\theta,k)}{(1-k^{2})\sin^{3}\theta}\rp, 
\end{eqnarray} 
with $F(\theta,k)$ and $E(\theta,k)$ are the incomplete elliptic integrals  
\begin{eqnarray} 
\label{elliptic} 
E(\theta,k)=\int_{0}^{\theta}\sqrt{1-k^{2}\sin^{2}\phi}\,\dd \phi,\nonumber \\ F(\theta,k)=\int_{0}^{\theta}\frac{1}{\sqrt{1-k^{2}\sin^{2}\phi}}\,\dd \phi,\hspace{0.5cm} 
k\equiv \sqrt{\frac{1-q_{2}^{2}}{1-q_{3}^{2}}}. 
\end{eqnarray} 
with $\theta\equiv \arccos\lp q_{3}\rp$. Equation (\ref{geom}) is a complicated equation whose solution is  
a function $q_{2}=q_{2}(q_{3})$ representing the allowed ratios for configurations  
in virial equilibrium with major axis rotation. In figure \ref{bif2}  
we plot these solutions with and without cosmological constant.
Note that the Maclaurin ellipsoid ($q_{2}=1$) is always a solution of (\ref{geom}), for any value of $q_{3}$.  
Jacobi ellipsoids (triaxial case) are allowed for values of $q_{3}$ below the bifurcation  
point $q_{3}<q_{3}^{\rm bif}$. For $\Lambda=0$, the bifurcation point is located at $q_{3}^{\rm bif}=0.583$.  
On the other hand, with cosmological constant, we find the following relevant effects.  
First, the bifurcation point becomes a function of $\rho_{\rm vac}$.  
As long as we increase the value of $\zeta$, the bifurcation point grows.  
This means that the separation between Jacobi and Maclaurin takes place at lower eccentricities.  
For instance, for $\zeta=0.6$ we get $q_{3}^{\rm bif}\approx 0.658$.  
Another interesting effect  is that the range of allowed $q_{3}$ also changes.  
Jacobi ellipsoids configurations are now restricted to $q_{3}^{\rm min}<q_{3}<q_{3}^{\rm bif}$.  
Below $q_{3}^{\rm min}$, the only allowed solution is again the Maclaurin ellipsoid.  
For $\zeta=0.6$, the minimum value of $q_{3}$ is $q_{3}^{\rm min}=q_{2}\approx 0.24$.  
Notice the value of $\zeta$ is a time dependent quantity, which depends on the density, $\rho$, of the astrophysical object 
we are investigating as well as on the value of $\rvac$. Since we are interested in systems which have already reached an equilibrium 
configuration we consider $\zeta$ as measured today. Hence, $\rvac=\frac{7}{3}\rho_{\rm bg}$, where $\rho_{\rm bg}$ is the background matter density.
Taking into account the classical result from the spherical collapse model in an Einstein-de Sitter universe, 
then $\rho\approx200\rho_{\rm bg}$. This gives $\zeta\approx 0.023$. 
Of course, this value may grow or decrease depending on the lower or higher density of the astrophysical system we are studying. 

\begin{figure} 
\begin{center} 
\includegraphics[angle=270,width=6cm]{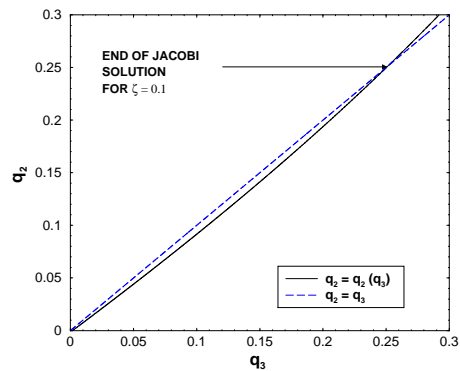} 
\end{center} 
\caption[Eccentricities II]{\footnotesize{Close plot for the second bifurcation point allowed for $\Lambda \neq 0$.}}\label{secbp} 
\end{figure} 
In figure \ref{secbp} we enlarged one plot for lower $q_3$ 
from figure 1 to explicitly show the second bifurcation point for  
$\zeta=0.1$. For $\zeta=0.06$, the second bifurcation point is located  
at $q_{3}\approx 0.12$, while for smaller values of $\zeta$, $q_{3}^{\rm min}$ goes to zero as expected.  
For practical purposes, the behavior of the bifurcation point $q_{3}^{\rm bif}$ can be approximately described  
as a function of $\zeta$ by a fit. We find in good approximation 
\begin{equation} 
 q_{3}^{\rm bif}(\zeta)=0.4826\lp 0.2082e^{8.207 \zeta}+1\rp, 
\end{equation}  
while the equilibrium condition can be fitted to be approximately 
\begin{equation} 
q_{2}(q_{3})=ae^{bq_{3}}-c, 
\end{equation}  
where the factors $a,b$ and $c$ (to be fitted) take different values as a  
function of the ratio $\zeta$.  
In table \ref{fittable} we summarize these values and our results on the bifurcation point  
of ellipsoids with major axis rotation.This bifurcation points separate triaxial solutions from oblate ones ($a_1=a_2$). 
\begin{table}
\begin{center} 
\begin{tabular}{|c|ccccc|}\hline \hline
$\zeta$         &$a$         &$b$         &$c$          &$q_{3}^{\rm bif}$ &$q_{3}^{\rm min}$\\ \hline  
$0$             &$0.34622$   &$2.3035$    &$0.334876$   &$0.582$           & $0$\\ 
$0.006$         &$0.329658$  &$2.3639$    &$0.314573$   &$0.585$           &$0.026$\\ 
$0.06$          &$0.332745$  &$2.3328$    &$0.318074$   &$0.589$           &$0.119$\\ 
$0.2$           &$0.348911$  &$1.9253$    &$0.334721$   &$0.657$           &$0.335$\\ 
\hline \hline
\end{tabular} 
\end{center} 
\caption[Fits]{\footnotesize{Fits for the equilibrium condition $q_{2}(q_{3})=ae^{bq_{3}}-c$  
for different values of $\zeta$ for the triaxial-oblate solution. The range of validity of Jacobi's solution is also shown.}} 
\label{fittable} 
\end{table} 
\begin{figure} 
\begin{center} 
\includegraphics[angle=0,width=4.cm]{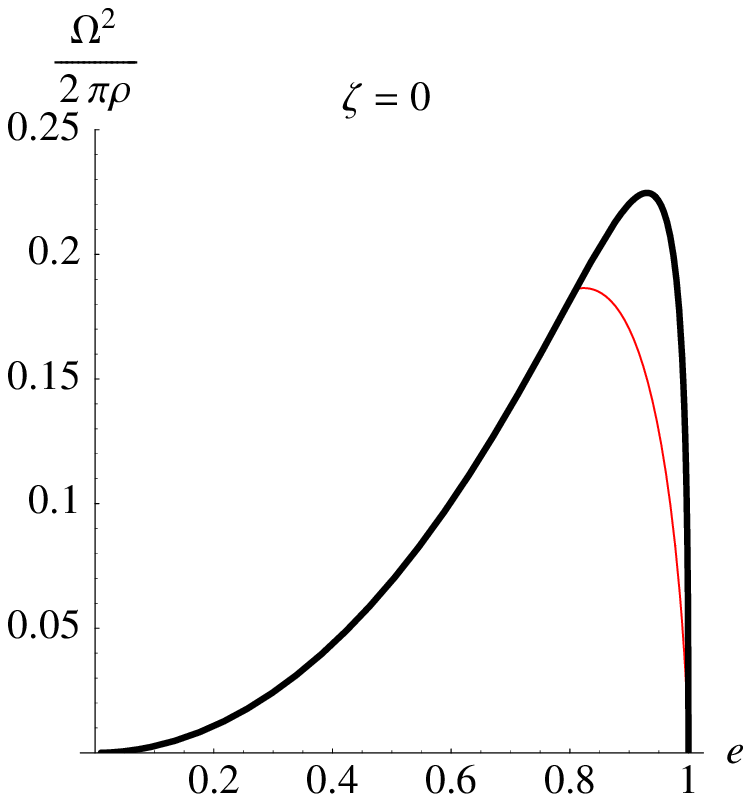} 
\includegraphics[angle=0,width=4.cm]{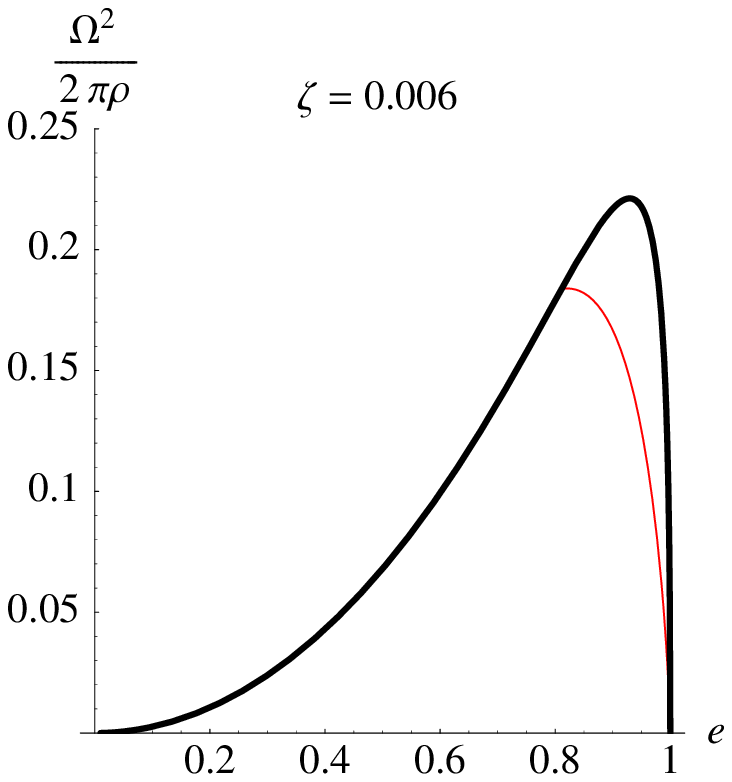}\\ 
\includegraphics[angle=0,width=4.cm]{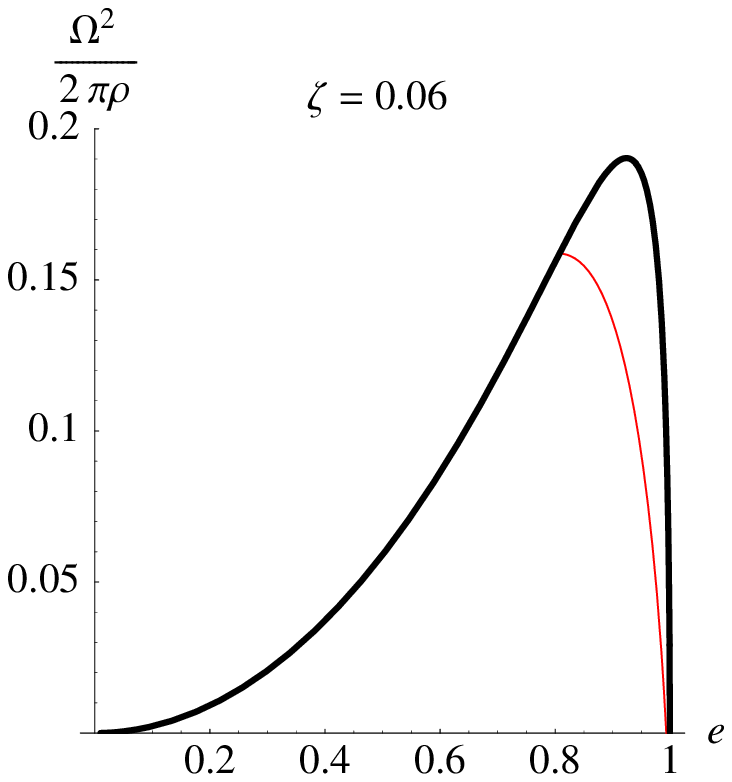} 
\includegraphics[angle=0,width=4.cm]{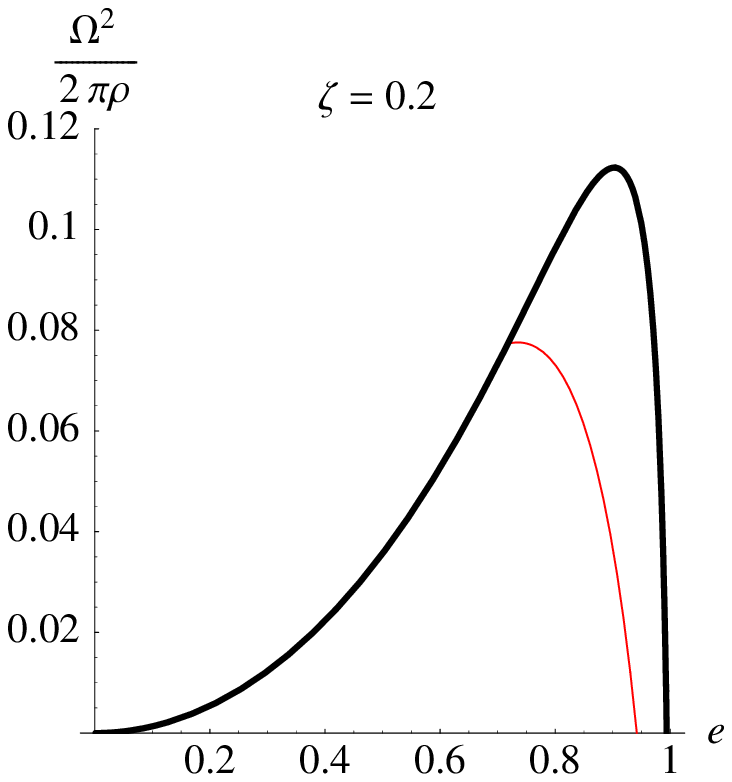}\\ 
\end{center} 
\caption[Maclaurin and Jacobi solutions for different ratios $\zeta$]{\footnotesize{Angular velocity  
of Maclaurin and Jacobi solutions for different ratios $\zeta$.  
The black line corresponds to the Maclaurin solution. The red line is the Jacobi solution.}}\label{ombif} 
\end{figure} 
\noindent Finally, in terms of the quantities $A_{i}$ defined at the beginning of this section, the angular velocity is written as 
\begin{equation} 
\label{bif} 
\frac{\Omega_{\rm rot}^{2}}{2\pi \rho}=A_{1}(q_{2},q_{3})-q_{3}^{2}A_{3}(q_{2},q_{3})-\frac{2}{3}\zeta\lp1- \frac{q_{3}^{2}}{q_{2}^{2}}\rp, 
\end{equation} 
In figure \ref{ombif} we plot the angular velocity against the eccentricity 
\begin{equation} 
e=\sqrt{1-q_3^{2}} 
\end{equation}  
The Maclaurin and Jacobi solutions are shown. The explicit dependence of $\zeta$ in the bifurcation point can also be seen.  
As in figure \ref{bif2}, one can infer from this plot the allowed ranges (in eccentricity) for the Jacobi's solution.  
One can also check the second effect of a non zero $\Lambda$ on rotating configurations,  
which is to reduce the angular velocity with growing value of $\zeta$ (see \cite{MNCh}).        

To make contact with real properties of galaxy clusters we note
that eccentricities of these objects have been extracted from simulations or observations
at different red-shifts $z$ \cite{floor, paz}. In some scenarios the distribution of clusters 
around eccentricity $1$ is non-negligible \cite{floor}. If these objects are triaxial then, however,
their eccentricity is limited to $e(q_3^{\rm min})$ according to table 1. On the other hand, the change in
$q_3^{\rm bif}$ will not be very significant.
In view of the many effects of $\Lambda$ on local astrophysical properties mentioned in the introduction,
this is however apriori not obvious.

\section{Minor axis rotation: triaxial and prolate ellipsoids} 
A note on convention is in order here. We define the prolate configuration by 
\begin{equation} 
a_1 > a_2=a_3 
\end{equation} 
to make a connection with the triaxial convention for semi-axis
(\ref{axis}) i.e we would like to call in both cases $a_1$ the major
axis.  This is in contrast to the convention $a_1=a_2 <a_3$ used in
\cite{Binney} which, given (\ref{axis}), is unsuitable to discuss a
connection between prolate and triaxial ellipsoids as we will do
below.  Note that if we use the function $A_i$ from reference
\cite{Binney} we have to interchange $A_1 \leftrightarrow A_3$ while
keeping the eccentricity to be $e^2=1-q_2^{2}$. 
The prolate geometry offers a better scenario to test the effects of a positive cosmological constant.  
This can be seen by considering a non rotating homogeneous configuration \cite{MNCh}.  
By solving for the mean velocity from the kinetic energy we can write \cite{MNCh}. 
\begin{equation}  
\label{mvel} 
\langle v^{2} \rangle =\frac{\rho^{2}|\tilde{\maw}^{\rm N}|}{2M}\left[1-\frac{1}{2}\maa \zeta\right]. 
\end{equation} 
This expression is valid for any geometry.  
The geometrical factor $\maa$  defined as $\maa \equiv (16\pi/3)(\tilde{\mai}/|\tilde{\maw})$  
where $\tilde{\maw}=2\rho^{-2}\maw$ and $\tilde{\mai}=\rho^{-1}\mai$,  
is what enhances the effects of $\Lambda$: in the limit of high eccentricities, one can determine 
$\maa_{\rm obl} 
\to  \frac{8}{3\pi} q_{3}^{-1}$ for oblate ($q_{3}\ll 1$) and  
$\maa_{\rm pro} 
 \to \frac{2}{3}q_{3}^{3}\left[\ln \lp 2q_{3}\rp\right]^{-1}$ for prolate ($q_{3}\gg 1$). 
The factor $q_{3}^{2}$ in found in the limit $a_{3}>a_{1}$ can enhance  
the effects of the density factor $\zeta$, in contrast to the oblate case,  
where the ratio is only found to the first power. It is then expected that minor axis rotation which is  
not possible for an oblate spheroid might have some large effects. 
 
 
Minor axis rotation is a rare case, since it is expected that the flatness  
of a configuration is in part due to the centrifugal forces acting on a rotating configurations,  
as is the case of oblate systems. Nevertheless,  
such behavior has been observed in galaxies \cite{minoraxis1} and we can therefore expect to find 
a similar behaviors in  
galaxy clusters which are known to rotate \cite{rot}.  
Whatever the dynamical reason for such a rotation, it is interesting to investigate this issue from
the equilibrium point of view. It has been stressed in the literature that a minor axis rotation is not possible
for oblate geometries. However, it was not realized that the triaxial case is not an equilibrium solution unless
we modify the virial equation, for instance, by the inclusion of $\Lambda$. 
 
The procedure of this section is similar to the previous section.  
We will explore the equilibrium condition and possible solutions in the case  
when the angular velocity lies along the major axis $\hat{e}_{\rm x}$,  
in accordance with the convention chosen in \cite{Binney} for the triaxial ellipsoid ($a_{1}>a_{2}>a_{3}$).  
The expression for the angular velocity is written in this case as 
\begin{equation} 
\label{omegapr} 
\Omega^{2}_{\rm rot}=\frac{|\maw_{zz}|-|\maw_{xx}|}{\mai_{zz}}-\frac{8}{3}\pi\rho_{\rm vac}\lp1 
-\frac{\mai_{xx}}{\mai_{zz}}\rp, 
\end{equation}  
while the geometrical restriction (\ref{gres}) becomes  
\begin{eqnarray} 
\label{gres1} 
\mai_{yy}\lp|\maw_{zz}|-|\maw_{xx}|\rp-\mai_{zz}\lp|\maw_{yy}| 
-|\maw_{xx}|\rp=\\\nonumber \frac{8}{3}\pi \rho_{\rm vac}\mai_{xx}\lp\mai_{zz}-\mai_{yy}\rp, 
\end{eqnarray} 
In terms of the functions $A_{i}$ defined in (\ref{a}), these expressions are written as 
\begin{eqnarray} 
\label{gres2} 
\frac{\Omega^{2}_{\rm rot}}{2\pi \rho}&=&A_{3}(q_{2},q_{3})-q_{3}^{-2}A_{1}(q_{2},q_{3}) 
-\frac{2}{3}\zeta\lp1-q_{3}^{-2}\rp \nonumber \\ 
q_{2}^{2}&=&\frac{q_{3}^{2}(A_{1}(q_{2},q_{3}) 
-\frac{2}{3}\zeta)}{A_{1}(q_{2},q_{3})+q_{3}^{2}(A_{2}(q_{2},q_{3})-A_{3}(q_{2},q_{3}))-\frac{2}{3}\zeta}. 
\end{eqnarray} 
Let us concentrate on the second line of Eq. (\ref{gres2}).  
Clearly, the prolate solution given in this case as $q_{2}=q_{3}$ is trivially satisfied  
in (\ref{gres1}) with $\mai_{zz}=\mai_{yy}$ and $\maw_{zz}=\maw_{yy}$, independent  
of the value of the ratio $\zeta$.  
One also notices that in the standard case $\Lambda=0$,  
the prolate solution  
is the only non trivial allowed figure in equilibrium with minor axis rotation.  
This implies that a triaxial configurations with minor axis rotation  
is not permissible as a configuration of equilibrium with zero cosmological constant.  
%

\begin{figure} 
\begin{center} 
\includegraphics[angle=270,width=8.cm]{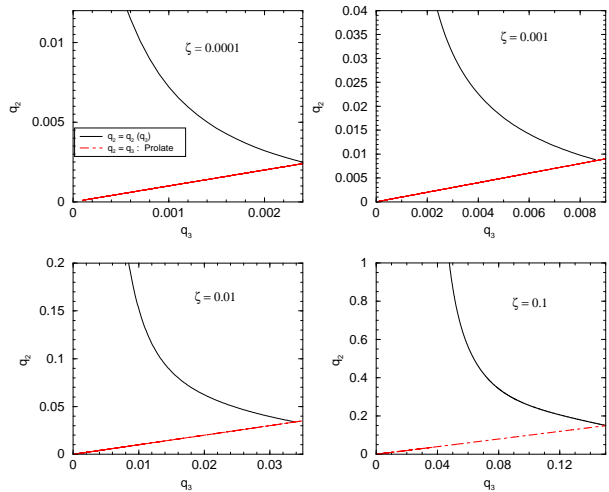} 
\end{center} 
\caption[Bifurcation point]{\footnotesize{Bifurcation Prolate-Triaxial:  
the line $q_{2}=q_{3}$ represents the prolate solution.}}\label{sol} 
\end{figure} 
%
Switching on a non-zero cosmological constant changes the situation.  
Surprisingly, for $\Lambda\neq 0$  
we find a bifurcation point from which a triaxial solution can be determined.  
This is shown in Fig. \ref{sol}, where we have plotted the solution $q_{2}(q_{3})$  
for different ratios $\zeta$. In this figure,  
the straight line represents the prolate solution $q_{3}=q_{2}$.  
The bifurcation point arises for small values of $q_{3}$,  
which represents large values in the eccentricity $e=\sqrt{1-q_{3}^{2}}$.  
For the largest value of $\zeta$,  $\zeta=0.1$,   
we obtain the largest value for the bifurcation point, located at $q_{3}\approx 0.15$.  
Hence the triaxial solution arises for $e>0.98$. Note that the triaxial solution  
has also a lower limit of validity for $q_{3}$. It occurs when $q_{2}$ approaches $1$.  
For instance, for $\zeta=0.1$, the triaxial solution is valid in the interval  
$0.047<q_{3}<0.15$ ($0.988<e<0.998$). As long as $\zeta$ decreases,  
the range of validity of the triaxial solution is reduced. The complete set of values  
of our solutions are given in table \ref{tablepro}. 
\begin{table} 
\begin{center} 
\begin{tabular}{|c|cc|}\hline  \hline  
$\zeta$         &$q_{3}^{\rm bif}$        &$q_{3}^{\rm min}$\\ \hline  
$0.0001$             &$0.0024$                 &$4.25\times 10^{-5}$ \\ 
$0.001$         &$0.00868$              &$4.25\times10^{-4}$ \\ 
$0.01$          &$0.0338$              &$0.00423$\\ 
$0.1$           &$0.15$              &$0.0478$  \\ 
\hline \hline  
\end{tabular} 
\end{center} 
\caption[Fits]{\footnotesize{The two bifurcation points for the transition  
triaxial-prolate for different values of $\zeta$. }} 
\label{tablepro} 
\end{table} 
%
\begin{figure} 
\begin{center} 
\includegraphics[angle=0,width=4.cm]{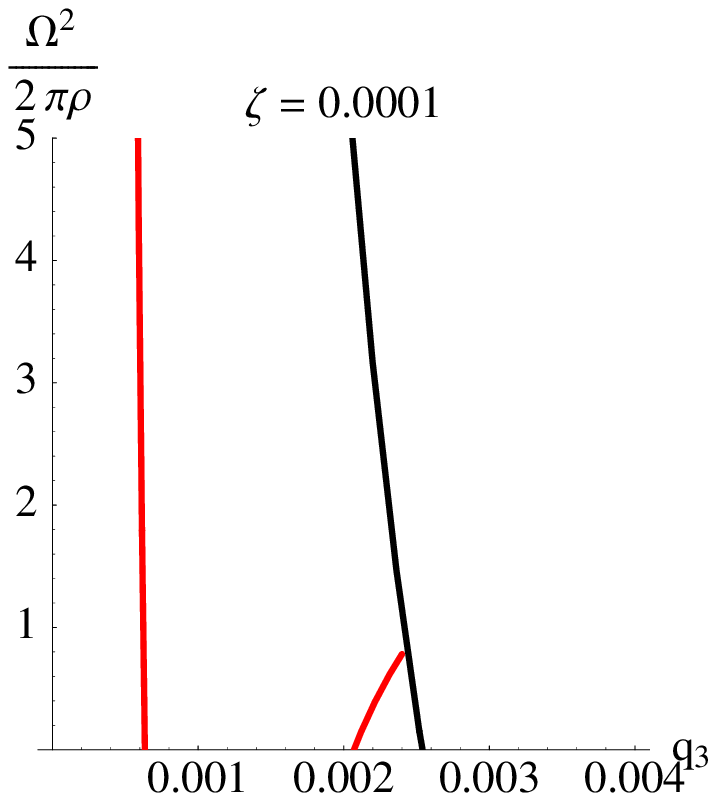} 
\includegraphics[angle=0,width=4.cm]{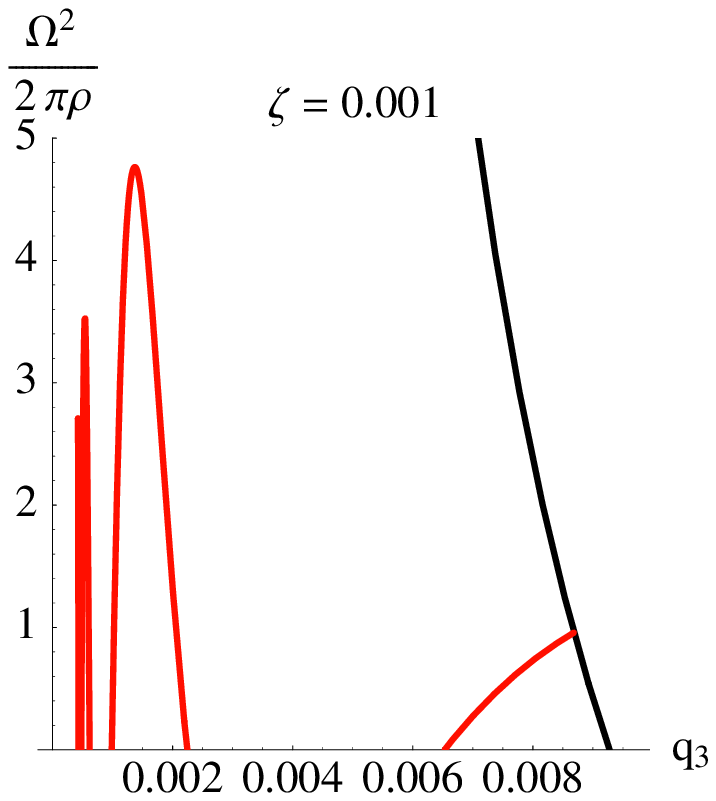}\\ 
\includegraphics[angle=0,width=4.cm]{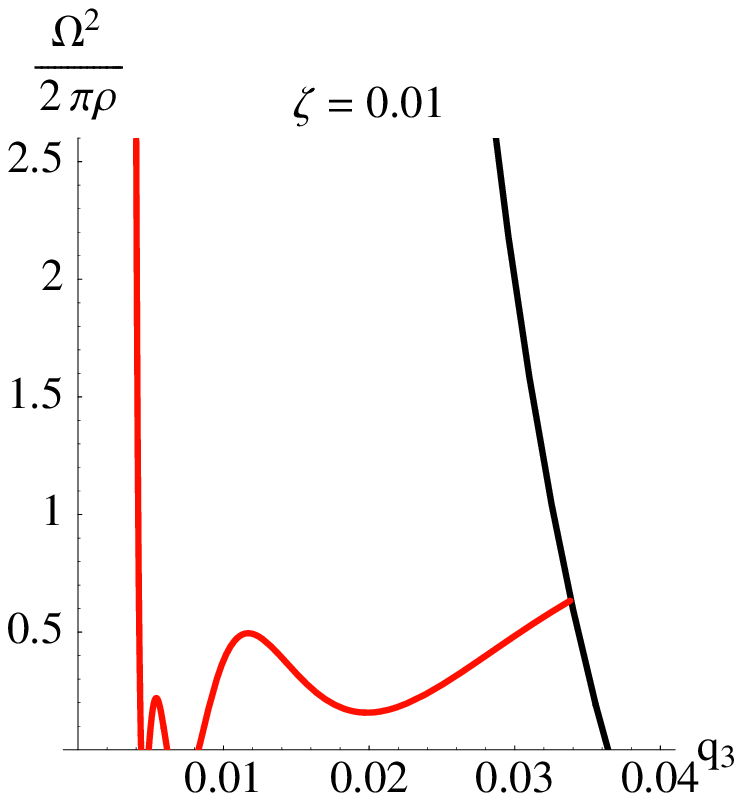} 
\includegraphics[angle=0,width=4.cm]{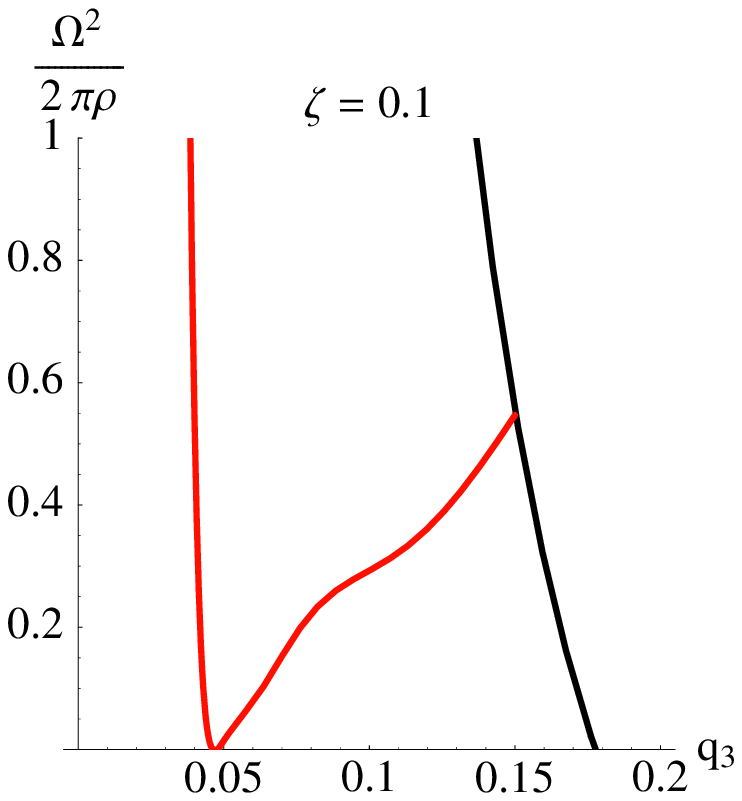}\\ 
\end{center} 
\caption[Maclaurin and Jacobi solutions for different ratios $\zeta$]{\footnotesize{Angular  
velocity of prolate and triaxial solutions for different ratios $\zeta$.  
The black line corresponds to the prolate solution .  
The red line is the triaxial solution allowed only with cosmological constant.}}\label{ompr} 
\end{figure} 
 
In figure \ref{ompr} we plotted the angular velocity against $q_3$. The bifurcation point, $q_3^{\rm bif}$,  
is where the prolate curve intersects the corresponding triaxial case. $q_3^{\rm min}$ is represented by the 
left vertical line beyond which we have no solution. It can be nicely seen that with decreasing $\zeta$ 
the range of the triaxial solution becomes smaller until it vanishes for $\zeta=0$.  
The angular velocity is also 
not a smooth function as it was the case for the triaxial/oblate case. Indeed, it oscillates (allowing for the  
moment the physically excluded range of $\Omega < 0$) when approaching $q_3^{\rm min}$.
As can be seen from figure \ref{ompr} the restriction of positivity on $\Omega$ narrows the allowed range
of $q_3$ further. Given the fact that without $\Lambda$ triaxial configurations with minor axis rotation are not
in equilibrium, this narrow range is rather a virtue than a disadvantage if we want to put the theory to test.
Had we an accurate knowledge of a triaxial ellipsoidal cluster rotating the minor axis, the theory
outlined above could be easily verified or falsified, of course bearing in mind the simplified assumptions of 
constant density/no internal motion (as in \cite{Chandra}) and the fact that only projected information of the axis is available to us.
From the observational side, its encouraging to know that dark matter halos around galaxy clusters are described
by triaxial configurations which tend to be more prolate \cite{paz}. 

\section{Conclusions}  

We have investigated effects of a positive cosmological constant on astrophysical objects with ellipsoidal geometries.  
We have considered both commonly known homogeneous ellipsoid configurations  
associated to cosmological and astrophysical systems, namely, oblate and triaxial configurations  
whose rotation is along the minor axis, as well as configurations with constant angular velocity along the minor axis. \\
\noindent We made use of the Newtonian limit of the Einstein field equations and have solved the tensor virial equation  
for the angular velocity. \\ 
\noindent Our results generalize the ones obtained in \cite{Chandra}  which applies for the case 
$\Lambda=0$. This is especially true for the 
bifurcation points of rotating ellipsoids as solutions of tensor virial equations for the oblate geometry. 
Adding to that we have also investigated the prolate/triaxial case with minor axis rotation,  
which has not been examined before as this case is only possible with $\Lambda \neq 0$. \\
\noindent We have found that the bifurcation point, which in the oblate case distinguishes 
the Maclaurin ellipsoid (oblate) from the Jacobi ellipsoid (triaxial), is sensitive to the cosmological constant.  
A new effect is given by the appearance of $q_3^{\rm min}$ which restricts the validity of the solutions from below. 
This point exists only for a positive cosmological constant. 
\noindent We also show that the cosmological constant allows triaxial configurations  
of equilibrium rotating along the minor axis as solutions of the virial equations.  
The significance of the result lies in the fact that minor axis rotation 
is indeed found in nature. Being impossible for the oblate case, it is permissible  
for prolate geometries, with $\Lambda$ zero and positive. 
However, for the triaxial case an equilibrium solution for the minor  
axis rotation is found only for non-zero positive $\Lambda$. 
Therefore the latter case is intimately, albeit indirectly,  related to the existence of a cosmological constant. 

The ellipsoidal geometry which we have studied 
in this paper is well suited to model many geometries, such as the disk. Therefore, the results of this article should in principle be 
applicable also to superclusters which are usually found to be very flat objects \cite{supercl}.  
 

\ack{DFM acknowledge support from the Research Council of Norway  
through project number 159637/V30. }


\section*{References}

\bibliographystyle{unsrt}

\end{document}